# The Origin of the Pseudogap in Underdoped HTSC.

Moshe Dayan

Department of Physics, Ben-Gurion University,
Beer-Sheva 84105, Israel.

E-mail: mdayan@bgu.ac.il

# The Origin of the Pseudogap in Underdoped HTSC.

Moshe Dayan

Department of Physics, Ben-Gurion University,
Beer-Sheva 84105, Israel.

## Abstract

Here the Origin of the pseudogap in HTSC is attributed to the modulated antiferromagnetic (AFM) phase, whose preliminary version has been sketched recently by the present author [1] ( J. Super. Nov. Mag. **22**, 517 (2009)). Starting from the t-J Hamiltonian, I show that the formal failure of the perturbation theory leads to a transformation to the pseudogap phase. This phase is characterized by the aggregation of the holes into rows and columns, which in turn results in two internal fields. The first is the modulated AFM field, whose main evidence comes from Neutron scattering experiments. The second internal field is made up by the checkerboard charge density waves that have been observed by Scanning Tunneling Measurements. The present paper deals mainly with the internal field of the first type, and discusses the second type only tentatively. Formalism is derived that yields the ground state, the internal field, the Hamiltonian, and the propagators of the condensed phase. Our results resolve the presumably inherent self contradictory concept of pseudogap. It is shown that the excitation energy spectrum is gapless despite the order parameter that is inherent to the condensed system. In addition, it is shown qualitatively that our model predicts "Fermi surface" that is in agreement with experiment.

## 1. INTRODUCTION

Despite the large volume of research on copper-oxide HTSC since its beginning, there is not yet one consensual general theory of the field. The experimental advancement, however, is impressive. It covers a broad range of experiments, from which we would like to focus on the following three: Angle Resolved Electron Photoemission Spectroscopy (AREPS), Neutron Scattering Measurements (NSM), and Scanning Tunneling Microscopy and Spectroscopy (STM and STS). These techniques have provided experimental evidence to some of the most intriguing features of these materials. Among these are the "Fermi surfaces" of these materials, which usually are made up by arks of metallic character in the nodal directions, and sections of non-metallic character in the antinodal directions. The surface section in each antinodal direction is made up by two lines that are symmetrical about the points $(\pm 1, 0)$, and $(0, \pm 1)$, (in units of $\pi a^{-1}$) in the Brillouin zone, and almost parallel near these points [2-11].

Another interesting observation by inelastic NSM is the appearance of a magnetic ordered field, at energies of the order of several meV to several tens of meV, which is centered on the wave-number $(1,1)$. This is the dynamic counterpart of the static AFM order in the undoped parent Mott insulators. At certain temperature ranges, and impurity concentrations, this magnetic peak appears to be modulated in the antinodal directions in the wavevector space, splitting into the four peaks $(1 \pm 2\delta, 1)$, and $(1, 1 \pm 2\delta)$, where $\delta$ represent incommensurability [12-16]. This kind of modulation has been observed in the impurity range x where the samples are conductive, namely for x>0.05. For smaller x, the modulation is in the nodal directions, and the observed peaks are at $(1 \pm 2\delta, 1 \pm 2\delta)$ [17]. It has been conjectured that the magnetic modulation is dynamically bounded by charge strips. In certain materials (most notably $La_{1.875}Ba_{0.125}CuO_4$, and $La_{1.48}Nd_{0.4}Sr_{0.12}CuO_4$ ), the lattice interaction with the charge and spin density waves reaches the point of lattice instability, and induces structural phase transformation, from low temperature orthorhombic (LTO) to low temperature tetragonal (LTT) [18-20].

Additional unusual phenomenon is the charge density modulation of a checkerboard geometry which has been observed by STM on the surfaces of several

HTSC [21-26]. In general, the periodicity of this checkerboard for the underdoped regime is four times the lattice period-4a, but larger periods (up to 5a or more) have been reported for larger doping. The checkerboard charge modulation of the 4a period is in the directions of the Cu-O bonds, namely in the antinodal directions. Modulation in the nodal directions has also been reported. The latter has been reported to have a positive dispersion with respect to the wavevector, contrary to the first [21, 24]. In Ref. [21, 24] the charge modulations were attributed to interference between quasiparticles on certain points on the Fermi surface, where the wavevector of the modulation equals the difference between the wavevectors of the quasiparticles. This is a reasonable model only when the said quasiparticles are on nesting sections on the Fermi surface. Otherwise, there should be a continuum of interfered wavevectors, and no modulation with defined periodicity is created. Consequently, only the charge modulation that fits the wavevector differences between the nested sections in the antinodal directions is in accord with this model.

Propositions of one dimensional charge and AFM solitons started short time after the experimental discovery of HTSC. Experimental observations which were interpreted this way date back two decades ago [12,27]. Theoretical works along this line also started not later [28], and continued till recently [29]. All these theoretical works were based on numerical calculations. Numerical analysis is a powerful tool in obtaining quantitative results that are often impossible to obtain by the regular analytical tools. However, it is usually a weaker tool for providing physical insight.

In this paper we shall show that all the experimental observations that are reported in former paragraphs are related to each other as causes and consequences. They all stem from the essential characteristic that the pseudogap phase is an ordered phase, where the positive charges of the holes are aggregated to form columns and rows that make up the borders between AFM regions of opposite phases. We show that this produces SDW with the wavenumber $2\pi a^{-1}\delta$ that modulate the fundamental AFM spin wave at $(1,1)$. These waves make up one of the internal fields that are associated with the pseudogap phase. This model explains the four satellite peaks at $(1\pm 2\delta,1)$ and $(1,1\pm 2\delta)$ in the results of NSM. We also show that this gives the Fermi surface its famous features, namely, arks in the nodal directions that are terminated by almost parallel (nested) sections by the zone boundaries.

The 4a period charge modulations in the directions of the Cu-O bonds result from the instability that is produced by the nesting in the antinodal directions, roughly according to the general model by the present author [30]. We shall show that this nesting, which is a feature of the Fermi surface, is a direct consequence of the two directional sections, where each represents a one-dimensional movement of the said AFM modulated wave. This charge modulation instability makes up the second internal field of the pseudogap phase. We show below that both internal fields are of checkerboard geometry, contrary to the general perception about the AFM modulated wave, which has been believed to be of strip geometry [18,20,29].

We start the present work with the assumption that the doping holes aggregate to form rows and columns, and show that this assumption agrees with experiment. Then we argue that the failure of the perturbation theory suggests a transformation to the pseudogap phase. We define columns and rows operators, suggest a ground state and Hamiltonian, and obtain the internal field, the excitation energy and the propagator. We show that, despite the condensation and the existence of order parameter, the excitation energy is gapless, justifying the concept- "pseudogap".

## 2. The AGGREGATION of the HOLES.

In the present paper superconductivity is not treated and we consider only the pseudogap phase. We assume conductive samples well in the underdoped regime. Samples in the extremely underdoped regime, x<0.05, which are not conductive (or superconductive), are not considered here because they are less investigated experimentally, and also because their relation to the above discussed effects is less obvious. The considered range of doping is widely treated by the following t-J Hamiltonian [32]

$$H_{tJ} = -\tau \sum_{<ij>s} (a_{is}^+ a_{js} + hc) + J \sum_{<ij>} (S_i \cdot S_j - \frac{1}{4} n_i n_j)$$
$$= H_t + H_J, \qquad (1)$$

where the quantities in Eq.(1) are defined as in [1], (except for the replacement of the $c_{js}$ operators by $a_{js}$). The Hamiltonians $H_t, H_J$ are the hopping part, magnetic part of the total Hamiltonian, respectively. The derivation of Eq. (1) was obtained in the reduced Hilbert space in which doubly occupied sites are excluded. The derivation also assumes that the exchange parameter $J$ is smaller than the hopping parameter $\tau$. Under this assumptions the undoped Mott insulator with exactly one electron per site is Néel AFM, with the un-renormalized ground state energy of $-JN^2$, where $N^2$ is the number of sites of the two dimensional lattice with $N$ sites in each dimension (and cyclic boundary conditions have been assumed). In the said AFM lattice one may define the magnetic periodicity units as $a(1,1)$, and $a(1,-1)$. This defines the AFM Brillouin zone (BZ) as diamond-like that is bordered by the four straight lines connecting the four points $(\pm 1, 0)$, and $(0, \pm 1)$ in the regular Brillouin zone. This AFM BZ is full, and each state in it contains the two spin states. Since this AFM zone is full, it may also be considered as the underlying (semi-conducting) Fermi surface. Any minimum excitation energy (the gap energy) involves a state on this surface. One should be puzzled, therefore, by the fact that a relatively small amount of doping (such as 5%) changes the character of the Fermi surface so drastically, from diamond-like to a surface composed of four arks terminated by nesting lines in the antinodal directions. In this paper we suggest that the Fermi surfaces of HTSC indicate the existence of one-dimensional characteristics (but in two directions). This intuition is supported by experiment. The satellite peaks observed by inelastic NSM were indeed interpreted to result from one dimensional AFM wave that is made up by two AFM regions opposite in phase.

A description of the above mentioned wave was given in [1], where it was argued that its main features stem from the aggregation of the holes into straight rows and columns. Here, we wish to make some modifications to that preliminary description, and also to some common perceptions in the literature. Let us for a start assume one dimensional stripe model such as the one depicted in Fig. 1b of [1]. It is a one-dimensional wave, and therefore, it is related to one-dimensional reciprocal zone. Let us ignore, the spin glass region in the doping scale, and assume that the pseudogap region starts from very minute doping, such as $x = 2N^{-1}$. This corresponds to two columns of holes, and (*N*-2) columns of spins, where the columns of holes separate between AFM regions of different magnetic phases, and move in the x-direction. This one dimensionality implies the replacement of the directionality of the unit cell and Fermi surface, from diagonal into square in the x and y directions. Thus, if we chose a square unit cell of double lattice parameter, the resultant first BZ would be the square whose corners are $\pi a^{-1}(\pm 0.5, \pm 0.5)$. This zone, however, can accommodate only half of the electrons. The other half occupies a second zone, and the x-dimensionality implies that the second zone should be in the x-direction. This double zone accommodates all the spins and we consider it to be the underlying (pseudo-gapped phase) Fermi surface. Within this surface there are (N-2) full columns of two spin states, and 2 empty columns at the edges of the zone, at $k_x = \pm \pi/a$. These edge wave numbers are a direct consequence of the one lattice parameter hopping Hamiltonian. The vertical size of this surface, though, is only one half of the corresponding size of the regular zone- namely, $-\pi/2a \leq k_y \leq \pi/2a$. Notice that our definition of the BZ is different from the definitions used in the numerical analyses of the charged magnetic stripes [28,29]. The BZ defined here is an electronic zone to be partially populated by one-dimensional dynamic q-states, which do not produce CDW or SDW, but are one dimensional itinerant states that are smeared over the whole volume. The charged (which eventually become not charged) SDW show up only in the condensed phase of section 3. As the doping increases, the number of empty columns in the surface increases, on the account of the full columns. Then, the empty part of the zone fits the width of the BZ that was exercised in numerical analyses.

In addition to the above description, and as depicted in Fig. 1a in [1], there are equivalent spatial zones in which the holes are arranged horizontally and consequently, the surface in the reciprocal space is halved in the other dimension. Therefore, assuming this model, one should find these two surfaces superimposed, where each part is related to a different spatial zone, and different orientation of holes. The two superimposed surfaces, being related to different spatial areas, are independent of each other. This is very different from the Fermi surfaces observed by ARPES. It lakes the ark-like sections of the Fermi surface, and also the widths in the antinodal directions are far too large to fit those observed by ARPES. The controversy with ARPES observations should be resolved by rejecting the stripe geometry and assuming the checkerboard geometry for the condensation of the holes. Thus, the given $\delta N^2$ holes (for $x = \delta$), should be divided into two halves, where one half is arranged vertically, and the other half-horizontally. With this new perception the width of the Fermi surface in the antinodal directions cannot stay equal to one half of the regular BZ, because it would contain too many electron states. With the checkerboard geometry, the same electronic states take part in two semi-one-dimensional movements, in the vertical and in the horizontal directions. Thus, the width becomes only one quarter of the regular BZ (between $-\pi/4a$ to $\pi/4a$ in each direction), since the area that was related to one direction in the stripe geometry is now divided between two directions. Moreover, now the two superimposed stripes in the BZ have a common area (of $\pi^2/4a^2$) which prohibitively represents 4 electrons per state. We shift this excessive area out of the crossed stripes area as depicted in Fig. 1a. Now the Fermi surface has acquired additional diagonal sections which may be considered to be the simplistic substitute of the arks in a more realistic model, or in the experimental results. The Fermi area in the center of the BZ (that is roughly bordered by the diagonal lines in the nodal direction) includes states that participate in movements in both directions- x and y. The areas in the anti-nodal zones include states with movements that are more one-dimensional. For comparison we show in Fig. 1b the ARPES results of Yoshida et al. [7] for the underdoped sample of LSCO with x=0.07. In general there is a good agreement with our model. In the anti-nodal directions the experimental data seem to be spaced closer than the prediction of our model, namely $\frac{1}{4}\frac{2\pi}{a}$. The reason for this will be clear in section 3, where we show that the occupation probability of the states is smeared around what is shown in Fig.1a. This smearing occurs both in the transversal and in the longitudinal directions, thus leaving smaller intensities for the states that mark the underlying Fermi surface in Fig.1a. Yoshida et al. reported such low intensities in these regions of the BZ that forced them to obtain the "Fermi surface crossing" only by interpolation from higher intensity regions [7]. We assume that these low intensities result from the said smearing, which causes the large experimental error which, in turn, causes the deviation from our model.

 The Fermi arrangement in Fig. 1a should not be considered as a Fermi surface simply because the states within it are not states of a Fermi liquid, and specifically for the following reasons. They are magnetically correlated as shown in Ref. 36. The area enclosed by the lines in the reciprocal space in Fig. 1a contains magnetically correlated spins, whereas the shaded part of it contains empty states that transformed to be the linearly aggregated holes. Apart from the magnetic correlations, the electronic states in the anti-nodal regions of Fig. 1a are unusual in the sense that they are virtually dispersion-less. This feature may be deduced from the forthcoming Eq. (18), which suggests that the dispersion of the aggregated holes is given by $\varepsilon_q$. This implies that the longitudinal dispersion of each individual spin in the anti-nodal sections is given by $\varepsilon_q/N$ - which is virtually dispersion-less. Then the states do not accumulate in the shown Fermi area according to increasing energy. Instead, the area shown in Fig.1a should be viewed as a kind of BZ that accommodates all the spin states that take part in the coordinated movements of the aggregated holes. The states of the aggregated holes, though, have the dispersive energy $\varepsilon_q$. They naturally "consume" the shaded states in Fig.1a because: 1) Their constituents are anti-particles (holes). 2) Their movements stem from steps of one site hopping of the spins. 3) Their location is in accord with the treatment of Section 3, and consequently, is in accord with NSM.

The minimum energy ARPES by the $(\pm \pi, 0)$ anti-nodal regions probe electrons that are adjacent to hole-aggregations that move in the $\pm x$-directions (and far from those that move in the y-direction). A minimum magnetic energy of $3J/2$ is needed for this photo-emission. As shown above, the energy is dispersion-less. This has been proven by many ARPES experiments [4,7,8,11]. The dispersion in the nodal direction is not negligible, especially for samples close to the optimal doping. The nodal directions dispersion cannot be understood by our simple model. Further investigations of more realistic models are needed for this.

The underlying "Fermi surface" suggested by Fig.1a, besides being similar to ARPES results, may account for the checkerboard charge modulation of the 4a period. The width of the "Fermi surface" in the antinodal directions, which is $2\pi a^{-1}/4$, fits exactly the spatial period of 4a of the charge modulation observed by STM. These CDW may in turn produce the physical barriers that confine the states to the quarter BZ width, since they usually are associated with some energy gap. In this respect, a relevant question is whether the electronic CDW is static or dynamic. The first alternative is usually accompanied by a soft phonon of the same period.

The same checkerboard geometry should apply also to the SDW, because they both stem from the same Fermi surface. To see how this comes about we assume that half of the doped holes- $\frac{1}{2}\delta N^2$ accumulate in rows, and the other half accumulate in columns. However, since a column of holes still occupies $N$ spatial sites, it should "consume" two columns of states in the Fermi surface. This is so because the number of states within two columns in the Fermi surface of the checkerboard geometry is equal to the number of states of one column in the Fermi surface of the stripe geometry. Therefore, an immediate calculation yields for the width of the holes section in the repeated BZ the value $2k_F = 2\pi a^{-1}\delta$, as is shown in Fig. 1a. Since this is exactly the width for the stripe geometry, one concludes that this width itself is not a valid criterion for the proper geometry. In the present work, the criterion is the shape of the Fermi surface, and its consequential checkerboard 4a modulation.

After we have established that the aggregated holes are arranged and move according to a checkerboard geometry, some questions should be addressed about the essential physics of this system. It is clear that the system has condensed to a different phase with a different symmetry. Condensed phases are usually characterized by ground states of different symmetry, by internal fields that preserve that symmetry in a self consistent manner, by an order parameter, and by an effective Hamiltonian that is consistent with all the above. The transformation to the condensed phase is characterized by choosing some selective state and favoring it over the many states that are statistically dominant in the uncondensed phase.

Although the pseudogap phase is characterized by the checkerboard geometry, we shall assume, **only for pedagogical purposes**, stripe geometry as a starting framework. At the end, the results will be presented according to the checkerboard geometry. Energy considerations in the underdoping regime suggest that the system should preserve the AFM order as much as possible despite the hopping part of the Hamiltonian $H_t$, which acts against this tendency. When the locations of the holes are independent of each other, magnetic order is destroyed by the movement of the holes. To maintain the said dynamic AFM order, the ground state should be composed of combinations of holes that are aggregated into columns and rows. To get an insight to the problem we resort to perturbation theory, but apply it in an unusual manner. Perturbation theory usually divides the Hamiltonian into $H = H_0 + H_{int}$, where $H_0$ is the Hamiltonian of the free particles, which is soluble, and $H_{int}$ is the interaction Hamiltonian which acts as a perturbation. In our case the magnetic interaction Hamiltonian - $H_J$ is the one whose solutions are known for any spatial configuration of spins, whereas $H_t$ is the one that perturbs the AFM order. Of-course we suspect that this method is inadequate, since it is usually assumed that $\tau > J$, which indicates that the perturbation serious does not converge. However, since in the underdoped regime the magnetic coupling applies to much more sites than the hopping parameter (which applies only to holes), the system reduces its ground state energy by preserving dynamic AFM. We shall argue that the breakdown of the perturbation theory is an indication of symmetry breaking, and transformation to the pseudogap phase. The system choice of this condensed phase translates into choosing one term of the perturbation serious, and avoiding all the others. This one term is of the order N, in the perturbation dimensionless parameter $\left|\tau/J\right|$, and normalized accordingly. We define the "kinetic picture" to be the equivalent of the interaction picture in the usual perturbation theory. Thus, we define the kinetic state vector as $|\phi(t)>= \exp(iH_J t)|\phi_S(t)>$, where $|\phi_S(t)>$ is the state in the Schrodinger picture. The kinetic Hamiltonian in the kinetic picture should be: $H_t(t) = \exp(iH_J t) H_t \exp(-iH_J t)$. In the kinetic picture, both the state vector and the Hamiltonian are time dependent, and they obey the equation

$$i\frac{\partial}{\partial t}|\phi(t)> = H_t(t)|\phi(t)> . \qquad (2)$$

In an equivalent manner to the usual perturbation theory, we find that the time dependence of the electronic creation and annihilation operators is

$$a_{js}^+(t) = a_{js}^+ \exp(-iJ\frac{n_{js}}{2}t), \qquad (3a)$$

$$a_{js}(t) = a_{js} \exp(iJ\frac{n_{js}}{2}t), \qquad (3b)$$

where $n_{js}$ is the number of nearest neighbors electrons of opposite spin to the one in the site j. When the perturbation is switched on adiabatically the Hamilton is usually written as $H(t) = H_J + e^{-\varepsilon|t|}H_t(t)$, where $\varepsilon$ is a positive small parameter, which eventually is made to approach zero [33]. This way the Hamiltonian coincides with the unperturbed Hamiltonian at very large absolute values of time, $|t| \to \infty$, and with $H_{tJ}(t)$ at small times, when $|t| << \varepsilon^{-1}$. Then, the state vector changes according to the time development operator

$$|\phi(t)> = CU(t,-\infty)|\phi(-\infty)> = CU(t,-\infty)|\phi_0>, \qquad (4)$$

where $C = |<\phi_0|U(\infty,t)U(t,-\infty)|\phi_0>|^{-1/2}$ is a normalization number, and $|\phi_0>$ is an eigenstate of the unperturbed Hamiltonian. The time development operator is

$$U(t,-\infty) = \sum_{n=0}^{\infty}(-i)^n \int_{-\infty}^{t} dt_1 \int_{-\infty}^{t_1} dt_2 ... \int_{-\infty}^{t_{n-1}} dt_n H_t(t_1)...H_t(t_n)e^{-\varepsilon(|t_1|+..+|t_n|)} \qquad (5)$$

According to Gell-Mann and Low theorem $|\phi(t)>$ is an eigenstate of $H_t$ [33], with the expectation value

$$E_t = <\phi(t)|H_t|\phi(t)>. \qquad (6)$$

A comparison between $U(t,-\infty)|\phi_0>$ and $H_t U(t,-\infty)|\phi_0>$ by means of Equations (4-6) suggests that any n-th term in the first sum equals the suitable (n-1)-th term of the second, when the first is multiplied by $E_t = (\frac{J}{2}\bar{n}_{js}) = \bar{J}$, where $\bar{n}_{js}$ is an averaged value of $n_{js}$. Since $\bar{n}_{js}$ is of order 2, $\bar{J}$ is of order J. Because the n-th term of $U(t,-\infty)|\phi_0>$ is of order $(\tau/\bar{J})^n >> 1$, larger n terms dominate, and divergence is avoided only because of equally diverging normalization parameter. Therefore, we get

$$H_t(t)|\phi(t)> = \bar{J}|\phi(t)>. \qquad (7)$$

Interestingly, the eigenvalue of the hoping Hamiltonian scales with the magnetic coupling rather than with the hopping parameter.

So far we have not defined $|\phi_0>$. Let us start by defining the column state $\phi_{qC}$ as

$$\phi_{qC} = C_q^+ |0> = (N-n)^{-1/2} \sum_{j=1}^{N-n} C_j^+ e^{-iqaj} |0>, \qquad (8)$$

where $|0>$ is the AFM ground state of ($N$-n) columns of spins, and $C_j^+$ creates a column of holes in the j-th column, while moving all the spins of larger indexed columns one site up. The column hole destruction operator $C_j$ annihilates the operation of $C_j^+$, so that $C_j C_j^+ = 1$. We also write $C_j |0> = 0$, for obvious reasons. The state $|\phi_{0C}>$ is defined as

$$|\phi_{0C}> = \{\prod_{q=\pi a^{-1}(1-\delta)}^{\pi a^{-1}(1+\delta)} C_q^+\} |0>. \qquad (9)$$

From the definition of the operators $C_j^+$ and their conjugates we find

$$\{C_i, C_j^+\} = \delta_{ij}, \qquad (10a)$$

and

$$\{C_i^+, C_j^+\} = 0. \qquad (10b)$$

Eq.(10b) is not obvious. It is valid only with the definition of the right phase according to

$$C_{i_1}^+ C_{i_2}^+ .... C_{i_n}^+ |0> = (-1)^p C_{j_1}^+ C_{j_2}^+ .... C_{j_n}^+ |0>. \qquad (11)$$

In Eq.(11), $(i_1...i_n)$ is a permutation of $(j_1...j_n)$, which is produced by p exchanges. Eqs.(10) and (11) are guarantied by choosing an odd number of holes per column. This is so because each row component of $C_j^+ C_i^+$ is actually composed of many $a_{ml}^+$ and $a_{ml}$ operators. The number of these operators differs by one depending whether $i > j$, or $i < j$. Consequently, Eqs.(10) deal with anti-commutators only when the number of spins per column is odd. Otherwise, one should deal with commutators. Notice, however, that an even number of spins still does not make $C_j^+$ a regular Bose operator, since a multiple occupation is forbidden. The problem is a collective excitation problem within the Fermi systems (like the problems of Cooper-pairs, plasmons, magnetons, etc). These problems are usually treated within Fermi systems, but the final result (such as the propagator) is Boson-like. The choice of odd number of spins in a column, when this number is huge should not affect the results. Eqs.(10) yield immediately the anti-commutation relations

$$\{C_k, C_{k'}^+\} = \delta_{k,k'}. \tag{12a}$$

$$\{C_k^+, C_{k'}^+\} = 0. \tag{12b}$$

Eqs.(9-11) include products such as $C_{j_1}^+ C_{j_2}^+ .... C_{j_n}^+ |0>$ whose definition needs some clarification. Each $C_j^+$ increases the number of columns by one, so that n such creation operators span the column index from one to $N$. When writing a product of creation operators we adopt the convention to use a plus sign (in Eq.(11)) when the operators are arranged with increasing column indexes from right to left, so that the operator with the lowest $j$ index is the first to operate on $|0>$. This way, the $i$-th operator (from the right) creates a column of holes at the $j$-th column, while moving the last column index, from $N-n+i-1$ to $N-n+i$.

We notice that each $\phi_{qC}$ is generated as a combination of states, each with one column of holes, and $|\phi_{0C}>$ is a combination of states each including n columns of holes. The state $|\phi(t)>$, however, is much more complex because it is obtained by the repeated application of the hoping Hamiltonian on $|\phi_{0C}>$, which destroys the order of the holes. Consequently, the state $|\phi(t)>$ cannot be expressed as a combination of states with n columns of holes, as $|\phi_{0C}>$. Nevertheless, it is still an eigenstate of $H_t$. However, the state $|\phi(t)>$ is not suitable to be the basis for the ground state of the condensed pseudogap phase. As stated before, ordered phases are made through the

selection of special ordered configurations out of the whole repertoire. This selection, in the present case, is that the net operation of all the Hamiltonians $H_t(t_i)$, in a given term of Eq.(5), is to transfer all the spins on one side of any column of holes into that column, without violating the AFM order. The same is valid for the other side of the column. For this operation, the minimum order needed is $N$, but higher orders are also allowed provided that the end result is the same. To demonstrate the operation we examine a special simple term of $U(t,-\infty)$ that moves the spins in an ordered sequential manner, from one side of the empty column, into it. We denote that time development term by $U_C(t,-\infty)$, the index of the empty column by $j$, and the index of the row in which the transfer occurs by $l$. The ordered transfer guarantees that $n_{ls} = 2$, for every $l$ (including the first and the last ones when cyclic boundary condition is applied also to the magnetic coupling). Then we have

$$U_C(t,-\infty)\phi_{qC} =$$

$$= \frac{-\tau(-i)^N}{(N-n)^{1/2}} \int_{-\infty}^{t} dt_1 H_t(t_1)....\sum_{j=1}^{N} e^{-iqaj} \int_{-\infty}^{t_{N-1}} dt_N e^{-\varepsilon(|t_1|+..+|t_N|)} a_{j1,s}^+(t_N) a_{(j+1)1,s}(t_N) C_j^+ |0>. \quad (13)$$

After performing the $dt_N$ integration, we are left with

$$U_C(t,-\infty)\phi_{qC} =$$

$$= \frac{-\tau(-i)^{N-1}}{J(N-n)^{1/2}} \int_{-\infty}^{t} dt_1 H_t(t_1)....\sum_{j=1}^{N} e^{-iqaj} \int_{-\infty}^{t_{N-2}} dt_{N-1} e^{-\varepsilon(|t_1|+..+2|t_{N-1}|)} H_t(t_{N-1}) a_{j1,s}^+(t_{N-1}) a_{(j+1)1,s} C_j^+ |0>$$

(14)

The operator $a_{(j+1)1,s}$ is now time independent, and we have eliminated the infinitesimally small $\varepsilon$ in the denominator. After all time integrations, and after performing the same calculations for the transfer of the spins of the ($j$-1) column, we get

$$U_C(t,-\infty)\phi_{qC}(-\infty) = (\frac{-\tau}{J})^N (N-n)^{-1/2} \sum_{j=1}^{N} e^{-iqaj-iJt}(C_{j-1}^+ + C_{j+1}^+)|0>$$

$$= (\frac{-\tau}{J})^N 2\cos(qa)e^{-iJt}\phi_{qC}(-\infty) = \phi_{qC}(t)/C_N. \tag{15}$$

$C_N$ is a normalization number. Now we have

$$\phi_{qC}(t) = 2\cos(qa)e^{-iJt}\phi_{qC}(-\infty) = C_N U_C(t,-\infty)\phi_{qC}(-\infty). \tag{16}$$

Equations (7) and (16) suggest that we may define a column Hamiltonian as

$$H_{tC} = e^{iJt}C_N H_t U_C(t,-\infty). \tag{17}$$

$H_{tC}$ operates on $\phi_{qC}$, rather than on $\phi_{qC}(t)$. The multiplication by $e^{iJt}$ removes its time dependence, thus moving the treatment back to the Schrodinger picture.

$$H_{tC}\phi_{qC} = 2J\cos(qa)\phi_{qC} = \varepsilon_q \phi_{qC}. \tag{18}$$

Doping starts at $q = m\pi a^{-1}$, where $m = \pm(2l+1)$ are odd numbers that mark the zone border-lines in the extended BZ, and continues on both sides of these points, until $|q - m\pi a^{-1}| = \pi a^{-1}\delta$. The energy dispersion for creating a hole in the antinodal direction and in underdoped regime is, therefore, quite small: $\Delta E \cong 2J[\cos(\pi\delta) - \cos(\pi\delta - \Delta qa)] = J[2\pi\delta a|\Delta q| + (\Delta qa)^2]$. Small energy dispersions in the antinodal region have indeed been reported by several ARPES experiments [4, 5, 34].

The reader has noticed that our selection of $U_C$ is the simplest possible. Actually, it is preferable to select $U_C$ that is a linear combination of time development operators that move the same spins to the said column but at random order. There are $N!$ possible permutations for the order of such movements, each with the same probability. In such a scenario the parameter $J$ should be replaced by some averaged value $\bar{J}$, of the same order of magnitude. However, for the sake of simplicity, we keep using $J$ in the following. This last selection of $U_C$ is preferable since it treats all sites of the column equally, whereas in the former one the $H_t$ Hamiltonian acts only on the last site of the column. Anyway, one should keep in mind that $\phi_{qC}$ is a collective state of the whole column despite the fact that in each of its components, $H_t$ acts only on one spin.

The "Fermi surface" in Fig.1a has four "arms", but in the extended zone one can work with two arms- one in the x-direction, and the other one in the y-direction. Each momentum has a longitudinal component, and a transversal component. For convenience let us choose a notation in which the zero of the longitudinal momentum is shifted according to $k = q - sign(q)\pi a^{-1}$ (here we consider the first BZ). The transversal momentum remains with the same notation (q), and the same zero as before. In the present analysis we take the transversal momentum to be dispersion-less. This assumption is still doubtful, and the possibility of some dispersion for the transversal component of the momentum should be re-examined.

The operators $U_C$, and $H_{tC}$ are designed to operate only on columns of holes. Therefore, when $H_{tC}$ operates on $|\phi_{0C}>$, it yields

$$H_{tC} |\phi_{0c}> = \sum_{k=-\pi a^{-1}\delta}^{\pi a^{-1}\delta} e_k |\phi_{0c}>, \qquad (19)$$

with $e_k = -2J\cos(ak)$. Now we are ready to switch to checkerboard geometry. The Hamiltonian becomes $H_{tCR} = H_{tC} + H_{tR}$, where $H_{tR}$ operates only on rows of holes. The ground state of $H_{tCR}$ becomes

$$|\phi_0> = \{\prod_{k_c,k_r=-\pi a^{-1}\delta}^{\pi a^{-1}\delta} C^+_{k_c} C^+_{k_r}\}|0>. \qquad (20)$$

The operators $C^+_{k_r}$, and $C^+_{k_c}$ create rows of holes, and columns of holes, respectively, and they anti-commute with each other (and with each other conjugates). The Schrodinger equation is

$$H_{tCR}|\phi_0> = \sum_{k_c,k_r=-\pi a^{-1}\delta}^{\pi a^{-1}\delta} (e_{k_c}+e_{k_c})|\phi_0> . \tag{21}$$

3. THE PSEUDOGAP STATE.

The system that has been analyzed in section 2 is not the condensed system, in the pseudogap phase, despite the aggregation of its holes. The state $|\phi_0>$ is a state that includes the physical entities that make up the pseudogap phase, but it should be rearranged in order to make up the ground state of the pseudogap phase. It does not yield any of the internal fields that characterize condensed phases. We engineer such a ground state in accordance with [30,31], where nesting is considered to be a trigger of instability, which induces rearrangement of states in the ground state. In principal, in each of the four "arms" of the "Fermi surface" of Fig.1a, there are two kinds of nesting induced instabilities, one is longitudinal to the AFM wave, and the second is transversal to it. The second kind is the instability that produces the 4a period CDW. At present, there are still some uncertainties about this instability, which make one cautious about its incorporation into the theory. Moreover, former treatments about nesting induced pseudogap phases and their effect on superconductivity have not considered two different kinds of nesting in the same system [30,31]. Such treatment is a major challenge by itself, and needs a separate establishment. Consequently, in the following we ignore this transversal nesting and its effect, despite our belief of its existence. Thus, the ground state of the pseudogap phase is defined as

$$|\psi_0> = \prod_{k_c,k_r=-\pi a^{-1}\delta}^{\pi a^{-1}\delta} (v_{k_c}+w_{k_c}C^+_{\bar{k}_c}C_{k_c})(v_{k_r}+w_{k_r}C^+_{\bar{k}_r}C_{k_r})|\phi_0> = |\psi_{0c}\psi_{0r}> . \tag{22}$$

In Eq. (22), $\bar{k}_c, \bar{k}_r = k_c, k_r - sign(k_c,k_r)2k_F$, respectively, with $k_F = \pi a^{-1}\delta$. Notice that $\psi_{0c}$ and $\psi_{0r}$ operate on different branches of the "Fermi surface", and consequently they are normal to each other. Each also commutes (or anti-commute) with any operator that operates on the other branch. These features enable us to carry out calculations on each branch independently, which is pedagogically favorable, and will be practiced in the following.

The holes column density is found by calculating the expectation value of its number operator

$$\rho(aj) = <\psi_0 | C_{jc}^+ C_{jc} | \psi_0> \qquad (23)$$

For simplicity we shall omit the subscript index c (for columns) from operators and state vectors. When the number operator is furrier transformed, we get

$$\rho(aj) = N^{-1} < \prod_{k \neq k_1} \psi_k | \psi_{k_1}^+ \sum_{q,q'} C_q^+ C_{q'} \exp[iaj(q-q')] \psi_{k_1} | \prod_{k \neq k_1} \psi_k >$$

$$= N^{-1}[1 + 2 v_{k_1} w_{k_1} \cos(2 k_F a j)] + <\psi_{0,k \neq k_1} | C_j^+ C_j | \psi_{0,k \neq k_1}> \qquad (24)$$

Note that in the first term of Eq.(24) the spatial wave-number is $2k_F$, independent of $k_1$. Repeated calculations on the rest of the state yield

$$\rho(aj) = N^{-1}[n + \sum_k 2 v_k w_k \cos(2 k_F a j)] \qquad (25)$$

Eq.(25) includes two terms, one is an averaged density, which is space independent, whereas the other one is space dependent. The space dependence of the second term is a cosine of wavenumber $2\pi a^{-1} \delta$, exactly as the waves detected by NSM. The directions also fit, the x and y directions.

Note that the wave in Eq.(25) portrays both electronic CDW and SDW, that have the same space period. Based on elastic NSM on $La_{17/8} Ba_{1/8} CuO_4$ (and some replacements) in the LTT phase, it has been argued that the wavelength of the CDW should be one half of its associated SDW [18,20]. However, these NSM results are different since they do not probe the electronic charge, but the ionic charge. Besides, they are related to static modulations in the LTT phase, whereas our results are time dependent and related to the regular LTO phase of LSCO and other HTSC. Moreover, in the following paragraphs we suggest that the electronic CDW may be cancelled, leaving only a dynamic SDW.

We have obtained a charge density wave with the same periodicity as the magnetic one because our $C_j^+$ operators were defined so that a column of charges separates between the AFM phase A for column indexes smaller than $j$, and the different AFM phase B for column indexes larger than $j$. Therefore, a column of charges indicates an AFM step function (a step jump between phase A and B). This implies that when the $C_j^+$'s are linearly combined to form waves, then the charge wave should be the space derivative of the magnetic wave, which in turn implies the same periodicity for CDW and SDW.

The duality between the magnetic phases A and B suggests that there should be, a complementary magnetic counterpart to the one described by Eq.(25), in which the roles of A and B are exchanged. Thus, $|\phi_0>$ of Eq.(9) should be redefine as

$$|\phi_0> = \frac{1}{\sqrt{2}} \prod_{k=-\delta\pi a^{-1}}^{\delta\pi a^{-1}} (C_k^+ |0>_A + C_k^+ |0>_B) = |\phi_0>_A + |\phi_0>_B . \quad (26)$$

Keeping the simplified notation practice of one dimensionality (despite the assertion of checkerboard symmetry), the ground state of the condensed phase is redefined as

$$|\Psi_0> = \prod_{k=-\delta\pi a^{-1}}^{\delta\pi a^{-1}} [(v_k + w_k C_{\bar{k}}^+ C_k^+)|\phi_0>_A + (v_k - w_k C_{\bar{k}}^+ C_k^+)|\phi_0>_B]$$

$$= \prod_{k=-\delta\pi a^{-1}}^{\delta\pi a^{-1}} (v_k + w_k C_{\bar{k}}^+ C_k^+)|\phi_0>_A + \prod_{k=-\delta\pi a^{-1}}^{\delta\pi a^{-1}} (v_k - w_k C_{\bar{k}}^+ C_k^+)|\phi_0>_B = |\Psi_0>_A + |\Psi_0>_B$$

$$(27)$$

With these definitions we get two anti-phased internal waves,

$$\rho_A(aj) = {_A<}\Psi_0 | C_j^+ C_j |\Psi_0>_A = (2N)^{-1}[n + \sum_k 2v_k w_k \cos(2k_F aj)] \quad (28a)$$

$$\rho_B(aj) = {_B<}\Psi_0 | C_j^+ C_j |\Psi_0>_B = (2N)^{-1}[n - \sum_k 2v_k w_k \cos(2k_F aj)] \quad (28b)$$

Note that the two oscillating parts of these charge densities come in opposite signs and cancels each other when added. However, although the oscillating charge is cancelled, the magnetic wave is reinforced. This is so because the magnetic wave due to phase A is basically anti-phased with respect to the magnetic phase due to phase B, and the additional spatial phase shift (which resulted from the minus sign in Eq. (27)) brings it to magnetic phase enhancement. Thus, Eqs.(28) predict SDW of wavenumber $2\pi a^{-1}\delta$, which is in agreement with NSM. The minus sign of $w_k$ in Eq.(27) is the right choice of phase, because it causes the elimination of the charge wave, and the enhancements of the magnetic wave, both contribute to reduce the energy of the ground state.

$$\rho_e(ja) = \rho_A(ja) + \rho_B(ja) = n/N \tag{29}$$

$$\rho_m(ja) = N^{-1}\xi \sum_k 2v_k w_k \sin(2k_F ja). \tag{30}$$

In Eq.(30) $\xi$ represents a complete AFM configuration. The question of the accompanying static CDW of the LTT lattice of $La_{17/8}Ba_{1/8}CuO_4$ is still valid. We speculate that it portrays a soft phonon that is due to its interaction with the magnetic wave $\rho_m(ja)$ of Eq.(30). This is conceivable since the interaction of such an AFM wave with the lattice ions is probably indifferent to the magnetic phase (A or B), and indifferent to the sign of the sine function, a feature which should double the periodicity of its interaction.

The redefinition of the ground state in Eq.(27) is signified by reversing the sign of the $w_k$ in the part that is contributed by the magnetic phase B. Examination of [30,31] shows that the Nambu-like field (or simply the Nambu field) should not be affected by the redefinitions

$$\tilde{\Psi}_k = \frac{1}{\sqrt{2}}\{\begin{pmatrix} C_k \\ C_{\bar{k}} \end{pmatrix}_A + \begin{pmatrix} C_k \\ C_{\bar{k}} \end{pmatrix}_B\}. \tag{31}$$

It is obtained from the following combination of the elementary excitations

$$\tilde{\Psi}_k = \frac{1}{\sqrt{2}}\{\begin{pmatrix}-w_k\\v_k\end{pmatrix}\gamma_{kA} + \begin{pmatrix}v_k\\w_k\end{pmatrix}\eta_{kA}^+ + \begin{pmatrix}w_k\\v_k\end{pmatrix}\gamma_{kB} + \begin{pmatrix}v_k\\-w_k\end{pmatrix}\eta_{kB}^+\}, \tag{32}$$

where

$$\gamma_{kA} = (-w_k C_k + v_k C_{\bar{k}})_A = -\gamma_{\bar{k}A} \tag{33a}$$

$$\gamma_{kB} = (w_k C_k + v_k C_{\bar{k}})_B = \gamma_{\bar{k}B} \tag{33b}$$

$$\eta_{kA} = (v_k C_k^+ + w_k C_{\bar{k}}^+)_A = \eta_{\bar{k}A} \tag{33c}$$

$$\eta_{kB} = (v_k C_k^+ - w_k C_{\bar{k}}^+)_B = -\eta_{\bar{k}B} \tag{33d}$$

The non-diagonal elements of the matrix that pre-factorize the vector $\begin{pmatrix}C_k\\C_{\bar{k}}\end{pmatrix}_{A,B}$ vanish, turning it into the unity matrix. The unperturbed Hamiltonian density is calculated from $H_0 = i\tilde{\Psi}^+(x,t)\frac{d}{dt}\tilde{\Psi}(x,t)$, to yield

$$H_0 = \frac{1}{2}\sum_{k,i}[(E_k + \Lambda_{k,i})\gamma_{k,i}^+\gamma_{k,i} + (E_k - \Lambda_{k,i})\eta_{k,i}^+\eta_{k,i} - (E_k - \Lambda_{k,i})] \tag{34a}$$

$$H_0 = \frac{1}{2}\sum_{k,i}\{\varepsilon_k \tilde{\Psi}_k^+ \tau_3 \tilde{\Psi}_k + \Lambda_{k,i}[\tilde{\Psi}_k^+ \tilde{\Psi}_k + \tilde{\Psi}_k^+ \tau_1 \tilde{\Psi}_k]\} \tag{34b}$$

In Eqs.(34a,b), $i = A, B$, is an index which sums over the magnetic phases A and B, $\varepsilon_k = E_k(w_k^2 - v_k^2)$, $\Lambda_{kA} = -\Lambda_{kB} = -2w_k v_k E_k$, and $E_k^2 = \varepsilon_k^2 + \Lambda_k^2$, and we assume that $w_k$, and $v_k$ are real and positive. Obviously the unperturbed Hamiltonian is diagonal with respect to $\gamma_{k,i}$ and $\eta_{k,i}$, but not with respect to $C_k$ and $C_{\bar{k}}$. When one does not differentiates between phases A and B, the $\pm\Lambda_k$ that add to the energy parameters of $\gamma_k^+ \gamma_k$ and $\eta_k^+ \eta_k$, could be interpreted as a shift of the Fermi level [30]. Here however, it is no longer possible, since the shift for phase A is reversed in sign to the shift for phase B. We notice that $\gamma_{kB,A}^+ | \Psi_0 >$ have the same eigenvalues as $\eta_{kA,B}^+ | \Psi_0 >$, respectively. Moreover, the eigenvalues of $\gamma_{kA}^+ | \Psi_0 >$ and $\eta_{kB}^+ | \Psi_0 >$, at the Fermi momentum, equal zero. This is an important result since it guarantees that the material is an electrical conductor despite its order parameter $\Lambda_k$. This feature is what makes $\Lambda_k$, an energy pseudogap rather than an energy gap. The eigenvalues of $\gamma_{k_F B}^+ | \Psi_0 >$ and $\eta_{k_F A}^+ | \Psi_0 >$ however, are far away from the Fermi level, and are equal to $2\Lambda_{k_F}$. We are tempted to assert that they are manifested by signatures (hump structure in the tunneling differential conductivity) that were attributed to a larger gap [35]. However, this assertion is not certain unless one incorporates into the theory the CDW of 4a period, since the latter might also be manifested by large energy "signatures".

The former sections suggest that for energies within one $|\Lambda_k|$ from the Fermi energy, one should focus the treatment only on $\gamma_{k,A}$ and $\eta_{k,B}$. Terms of the field operator that are related to $\gamma_{k,B}$ and $\eta_{k,A}$ contribute only at higher energies. Thus, we redefine the field operator

$$\Psi_{k-}(t) = \begin{pmatrix} -w_k \\ v_k \end{pmatrix} \gamma_{kA} \exp(-iE_{k-}t) + \begin{pmatrix} v_k \\ -w_k \end{pmatrix} \eta_{kB}^+ \exp(iE_{k-}t) = \Psi_k(t). \quad (35a)$$

$$\Psi_{k+}(t) = \begin{pmatrix} w_k \\ v_k \end{pmatrix} \gamma_{kB} \exp(-iE_{k+}t) + \begin{pmatrix} v_k \\ w_k \end{pmatrix} \eta_{kA}^+ \exp(iE_{k+}t) \quad (35b)$$

Notice that these fields are no longer Nambu fields, since each is a sum of two terms that are related to two different phases (A and B), and the non-diagonal elements of their matrix do not vanish. The energies in Eqs.(35) are defined as

$$E_{k\pm} = \sqrt{\Lambda_k^2 + \varepsilon_k^2} \pm |\Lambda_k| \quad (36)$$

Hereafter we concentrate on the field $\Psi_{k_-}$ (which is simply denoted as $\Psi_k$), that is associated with the low excitation energy $E_{k_-}$. The unperturbed propagator $G_0(k,t) = -i <\psi_0 | T\{\Psi_k(t), \Psi_k^+(0)\} | \psi_0>$ is obtained as usual [30]. Its Furrier transform is

$$G_0(k,\omega) = \frac{1}{2}[\frac{I + (w_k^2 - v_k^2)\tau_3 - 2w_k v_k \tau_1}{\omega - E_{k_-} + i\delta\omega} + \frac{I - (w_k^2 - v_k^2)\tau_3 - 2w_k v_k \tau_1}{\omega + E_{k_-} - i\delta\omega}]. \tag{37}$$

The field operator obeys unusual anti-commutation relation

$$\{\Psi_k, \Psi_{k'}^+\} = (I - 2w_k v_k \tau_1)(\delta_{k,k'} I + \tau_1 \delta_{k,\bar{k}'}) \tag{38}$$

This relation suggests the existence of propagators that violate momentum conservation by $2k_F$, namely the propagators

$$G_0(k, \bar{k}, \omega) = G_0(k, \omega)\tau_1 , \tag{39a}$$

and

$$G_0(\bar{k}, k, \omega) = \tau_1 G_0(k, \omega) . \tag{39b}$$

In the field perturbation theory the condensation function $\Lambda_k$ is usually obtained by the Hartree and the Fock integrals, using the interaction Hamiltonian, which incorporates all possible interactions,. The interaction Hamiltonian for the present problem has never been worked out. Some problems are inherent to this Hamiltonian. First one has to formulate basic quantum interactions between columns of charges that are imbedded in the AFM matrix. How these interactions are integrated from the basic interactions of single particles (and anti-particles), provided that the final products remain columns or rows. A special question is about magnetic interactions, since the interacting species- $C_k^+ | \Psi_0 >_{A,B}$ are magnetic, besides being charged. Some progress has recently been achieved in this respect, and it is published in Eqs. (4) and (5) of Ref. 36. The contribution of the Hartree integral is critically dependent upon the matrix components of the interaction vertex [30,31,37]. It is shown in [37] that, by choosing the proper field, a vertex component proportional to $\tau_1$ is achieved, which in turn, facilitates the contribution of the Hartree diagram to $\Lambda_k$. The contributions of the Hartree diagram come from two wavevectors- zero and $2k_F$. We notice that while each aggregated state interacts magnetically only within its own phase (A or B), it interacts electrically with every other charge (weather it is in phase A or B). For the Hartree integrals of zero wavevector, the contribution of the Coulomb interactions that are screened by phonons and electron polarizations should yield zero, due to the required Coulomb stability. For $2k_F$, the Coulomb contribution to the Hartree integral is also zero, due to the anti-phase cancellation effect as in Eqs.(28a) and (28b). This conclusion is based on the understanding that the Hartree integral (for the $2k_F$ Coulomb interaction) gives the Coulomb interaction of a given propagator with the total charge of $2k_F$ wavenumber within the layer. The latter, though, is proven to be zero by Eq. (29). Thus, as far as the Hartree contributions, they result only from the positive contributions of magnetic interactions, and el-phonon-el interactions. The Fock contribution is more complex, but one feature of it is clear- it incorporates only interactions within the same magnetic phase. This is so for the same reason why Fock diagrams in regular metals connect only propagators of the same spin. We speculate that for $2k_F$, the Fock integrals should also be positive, due to the added positive magnetic interaction, and the estimation that the Coulomb interaction is screened by phonons and electron polarizations for such long ranges. Thus, the contributions of the Hartree diagram and the Fock diagram, should guarantees the existence of the pseudogap parameter.

## 4. CONCLUSION

Before concluding I wish to make some comments about the underlying "Fermi surface" in Fig. 1a. It has been obtained merely by the assumption of the linear aggregation of holes, which move in the x and y directions between regions of AFM ordered spins. It is based only on symmetry breaking, and has no adjustable parameters. This is contrary to the usual method for obtaining "Fermi surfaces" that fit the experimental ones. This method adds the term $t'\cos(k_x a)\cos(k_y a)$ to the band, where $t'$ is the second nearest neighbor hoping parameter. This hopping parameter is adjusted to get the concave curvatures that typify the "Fermi surfaces" of the cuprates. The ratio needed is quite large $|t'/\tau| \approx 0.3 - 0.4$ [38-41]. A direct evaluation of $t'$ suggests that it should vanish. This is simply based on the spatial symmetry of the $\sigma$-anti-bonding $Cud_{x^2-y^2} - O_{p_{x,y}}$ orbital, which has nodes along the lines connecting the second nearest neighboring Cu atoms. Thus, when attempting the integral for $t'$, integration with respect to the directions normal to the nodes has anti-symmetric integrands, which yield zero for the integral. Contrary to this obvious result, the calculations which obtain large $t'$ employ an intricate 8 orbitals scheme, with several indirect hopping channels, involving higher energy Cu-4s orbital, apical oxygens, $pd - \pi$ anti-bonding in addition to $pd - \sigma$ anti-bonding orbitals [39-41]. Such a multi-component complex soup is to be compared with the directly deduced Fig. 1a, which is solely based on symmetry breaking.

The present paper suggests that the origin of the pseudogap in underdoped Cu-oxide HTSC is the AFM waves (of checkerboard geometry) that have been observed by NSM. This structure, which is prevalent in underdoped samples, lowers the energy of the ground state by maintaining partial dynamic anti-ferromagnetism even when the material is doped by holes. Its existence is maintained by the aggregation of the holes into columns and rows. We have shown that this model is consistent with experiment, in particular with experimental Fermi surfaces, NSM results, and with STM results on the checkerboard CDW structures (of periods 4a).

We have demonstrated that the known t-J Hamiltonian could lead to the aggregation of holes, when the magnetic part of the Hamiltonian is assumed as the basic part and the kinetic part is assumed as a perturbation. After doing so, we have transferred the problem from two dimensions into a semi one dimensional problem. This contains a transformation from single particle operators to operators of rows and columns of particles. The Hamiltonian and the propagators were adjusted accordingly. The field operator is a two dimensional vector over the N-components row and column creation and annihilation operators. We have shown that, by using the two analogous phases of anti-ferromagnetism, we could achieve an energy reduction, and also prove the pseudogap character of the order parameter. The excitation spectrum is found to be gapless despite the existence of the basic features of condensed phases, such as an internal field and an order parameter.

In addition, we have shown qualitatively that by the mere assumption of aggregation of holes according to checkerboard geometry, the well known features of the Fermi surface may be obtained.

The formalism that has been formulated in the present paper yields internal fields, effective Hamiltonian and propagators. It also provides a formal way for calculating the order parameter. For practical calculations, we suggest further analysis of the relevant interactions between the rows/columns of holes that are imbedded in AFM matrix.

28REFERENCES

1. M. Dayan, J. Supercon. Nov. Magn. **22**, 517 (2009). See corrected version in arXiv:0901.3896v2. cond-mat. (2009).
2. H. Ding, et al, J. Phys. Chem. Solids **59**, 1888 (1998)
3. M. R. Norman, et al. Nature **392**, 157 (1998).
4. A. Ino et al. Phys. Rev. B **62**, 4137 (2000).
5. A. Ino et al. Phys. Rev. B **65**, 094504 (2002).
6. T. Yoshida et al. Phys. Rev. Let. **91**, 027001 (2003).
7. T. Yoshida et al. Phys. Rev. B74, 224510 (2006). See also arXiv: Cond. Mat./0510608v1 (2005).
8. T. Yoshida et al. J. Phys. Cond. Mat. **19**, 125209 (2007).
9. T. Kondo et al. Phys. Rev. Let. **98**, 267004 (2007).
10. K. Terashima et al. Phys. Rev. Let. **99,** 017003 (2007).
11. M. Shi et al. Phys. Rev. Let. **101,** 047002 (2008).
12. S. W. Cheong et al. Phys. Rev. Let. **67**, 1791 (1991).
13. T. R. Thurston et al. Phys. Rev. B, **46**, 9128 (1992).
14. K. Yamada et al. Phys. Rev. B, **57**, 6165 (1998).
15. N. B. Christensen et al. Phys. Rev. Let. **93**, 147002 (2004).
16. J. M. Tranquada et al. Phys. Rev. B, **69**, 174507 (2004).
17. S. Wakimoto et al. Phys. Rev. B, **63**, 172501 (2001).
18. J. M. Tranquada et al. Phys. Rev. B, **54**, 7489 (1996).
19. M. Fujita et al. Phys. Rev. Let. **88**, 167008 (2002).
20. M. Fujita et al. Phys. Rev. B, **70**, 104517 (2004).
21. J. E. Hoffman et al. Science **297**, 1148 (2002).
22. C. Howald et al. Phys. Rev. B, **67**, 014533 (2003).
23. T. Hanaguri et al. Nature **430**, 1001 (2004).
24. K. McElroy et al. Phys. Rev. Let. **94**, 197005 (2005).
25. Y. Kohsaka et al. Science **315**, 1380 (2007).
26. T. kurosawa et al. Phys. Rev. B, **81**, 094519 (2010).
27. R. J. Birgeneau et al. Phys. Rev. B, **39**, 2868 (1989).

FIGURE CAPTIONS.

Fig.1a

Proposed tentative Fermi surface, for underdoped high temperature Cu-Oxide Superconductor, is shown in two repeated BZ. The area marked by crisscross pattern represents states missing by hole-doping. Note that according to section 3 the state occupation, in the anti-nodal directions, becomes smeared.

Fig.1b

Experimental data of Yoshida et al. for LSCO with x=0.07 is compared with our model. The thick black lines represent the underlying Fermi surface of our model.

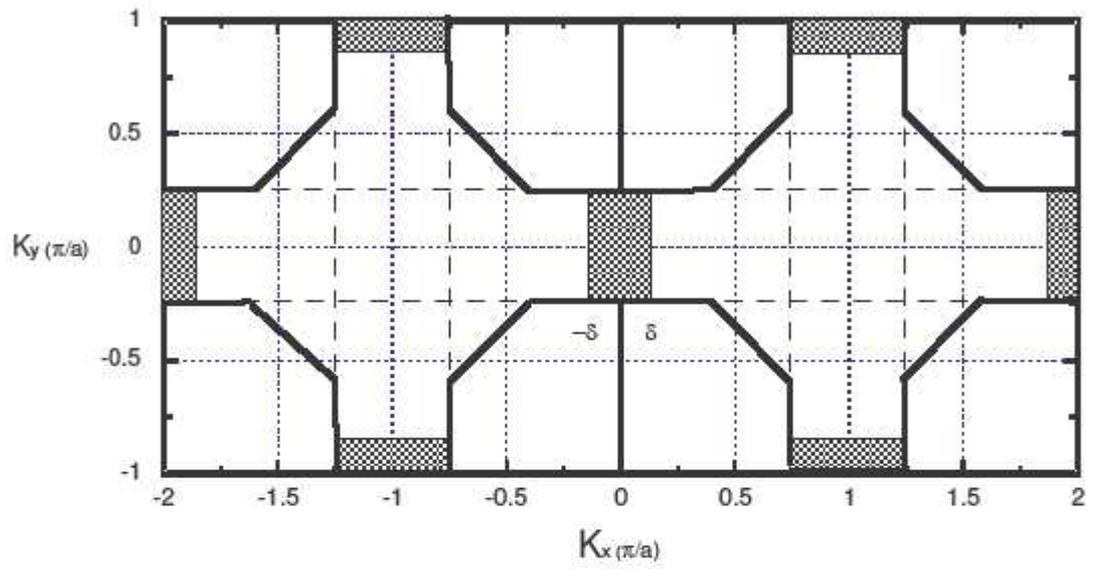

**Figure 1a**

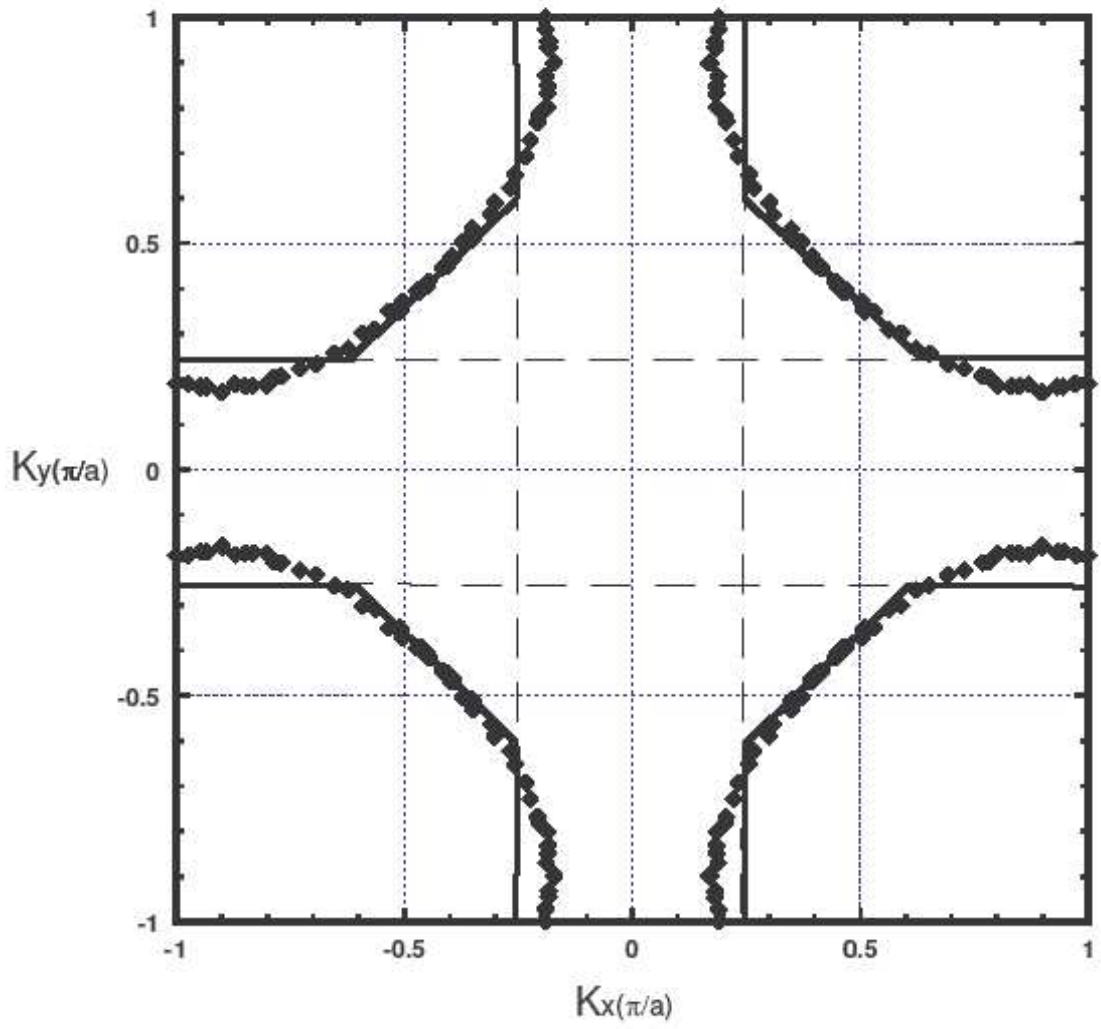

Figure 1b